\documentclass[aps,showpacs,amsmath,amssymb]{revtex4}

\usepackage{graphicx,epsf}
\usepackage[]{latexsym}
\newcommand{\be}{\begin{eqnarray}}
\newcommand{\ee}{\end{eqnarray}}

\begin{document}
\title{Dark Energy from Casimir Energy on Noncommutative Extra Dimensions}
\author{S. Fabi}
\email{fabi001@bama.ua.edu}
\affiliation{Department of Physics and Astronomy, The University of
Alabama, Box 870324, Tuscaloosa, AL 35487-0324, USA}
\author{B. Harms}
\email{bharms@bama.ua.edu}
\affiliation{Department of Physics and Astronomy, The University
of Alabama, Box 870324, Tuscaloosa, AL 35487-0324, USA}
\author{G. Karatheodoris}
\email{karat002@bama.ua.edu}
\affiliation{Department of Physics and Astronomy, The University
of Alabama, Box 870324, Tuscaloosa, AL 35487-0324, USA}
\begin{abstract}
We study the possibility that dark energy is a manifestation of the Casimir energy on extra dimensions with the topology of $S^2$. We consider
our universe to be $M^4 \times S^2$ and modify the geometry by introducing noncommutativity on the extra dimensions only, i.e. replacing $S^2$
with the fuzzy version $S_{F}^2$. We find the energy density as a function of the size of the representation $M+1$ of the algebra of $S_{F}^2$,
and we calculate its value for the $M+1=2$ case. The value of the energy density turns out to be positive, i.e. provides dark energy, and the
size of the extra dimensions agrees with the experimental limit.
We also recover the correct commutative limit as the noncommutative parameter goes to zero.%
\end{abstract}
\maketitle
\large
\section{Introduction}\label{intro}
The discovery that the expansion of the universe is currently accelerating \cite{riess,perlmutter,schmidt,blakeslee}
is now eight years old, and so far the cause of the phenomenon has not been rendered explainable in terms of conventional
classical or quantum physics.  The unknown cause of the acceleration has been given a label, dark energy, to facilitate
discussion of the effect.   Our inability to describe the phenomenon within the context of conventional physics suggests
that new physics is required for its explanation. Many explanations of dark energy in terms of more fundamental, and
usually as yet undetected, features of our universe have been put forward.  In this paper we show that acceleration of
the expansion of the universe can be obtained by combining an established fact of nature, the Casimir effect as calcuated
for a scalar field, with two
new (and so far undetected) physical properties of the universe --  large extra dimensions and noncommutativity of space.
\par
In this paper we  calculate the Casimir energy density of $M^{4}\times S_{F}^{2}$  following the method for large extra dimensions used by R.
Kantowski and A. Milton in~\cite{milton1}.  We  consider a massless scalar field and calculate the effect on the Casimir energy due to
noncommutativity (NC) on $S^2$. Noncommutativity is well understood for the  example $S_{F}^2$. The Green's function expansion in terms of
spherical harmonics is truncated to $M$, with $M+1$ being the size of the representation of the noncommutative algebra of
$S_{F}^2$~\cite{balac}. For this reason it becomes possible to perform calculations similar to those done in~\cite{milton1} and to deal with all
of the infinities in a similar way.

\par The energy density we obtain is of the form
 \be\label{U}
  u_{C}=\frac{\alpha_{TOT}}{a^{4}}\ln\frac{a}{b} \; ,
 \label{uc1}
 \ee
 where $\alpha_{TOT}=\alpha_{KM}+\alpha_{NC}+\alpha_{C}$, $a$ (the radius of $S^2$) and $b\simeq L_{Planck}$. The
 term $\alpha_{KM}$ is the one
 obtained in~\cite{milton1} while the other two terms are
 corrections due to noncommutativity, and they both depend on $M$.
Taking the $\lim {M \rightarrow \infty}$  (the commutative limit) the two noncommutative terms go to infinity and
are discarded in other approaches;
 Eq.(15) in~\cite{mde} is recovered after discarding certain infinities
\par It is worth describing this infinity juggling in plain language.
 In \cite{mde} the Casimir energy of $S^2$ is interpreted, from the point of view of an observer living on $M^4$,
 as an energy density filling this space with the appropriate equation of state to serve as dark energy.
 To obtain a finite result as in \cite{milton1} two thing are done: first only finite terms logarithmically dependent
 on $a$
 are retained and second a Planck scale cutoff is imposed.  In the noncommutative approach pursued here new finite
 terms proportional
 to $\ln(a/b)$ appear.  Of course, in the commutative limit these terms grow to infinity and reproduce exactly the
 terms thrown away
 in ~\cite{milton1}.
 Kantowski and Milton could have recovered a positive cosmological constant by adding another ad hoc energy cutoff
 (separate from $b$) but this procedure is terribly arbitrary.  Our approach is more natural for two reasons.
 Firstly, the ambiguity is reduced from the choice of a real number to the choice of a positive integer, and secondly this integer is related to
 the radius and noncommutativity parameter via (\ref{uno}).

\section{The Dark Energy Problem}

Observational evidence \cite{riess, perlmutter} from high-redshift
Type Ia supernovae indicates that the expansion of the universe is
accelerating.  The cause of the acceleration is unknown, but the
Robertson-Walker solution of general relativity requires dark energy to
be present in order for the universe to be accelerating. Many ideas and models have been proposed so
far~\cite{darkenergy} to describe the physical nature of dark
energy. Using the RW metric
 \be
 ds^2=g_{\mu\nu}dx^{\mu}dx^{\nu}=-dt^2+a(t)^2 \Big[ \frac{1}{1-\kappa r^2}dr^2+r^2d\theta^2+r^2\sin^2\theta
 d\varphi^2\Big]
 \ee
 with the energy-momentum tensor of a perfect fluid in the comoving frame
 \be
 T_{00}=\varrho  \qquad T_{ij}=p
 \ee
in Einstein's equation
 \be
 R_{\mu\nu}-\frac{1}{2}R\, g_{\mu\nu}= 8\pi G\, T_{\mu\nu}
 \ee
the two Friedman equations are
 \be
 \frac{\ddot{a}}{a} =-\frac{4\pi G}{3}\Big(\varrho+3p \Big)\\
 \Big( \frac{\dot{a}}{a}\Big)^2=\frac{8\pi G}{3}\varrho-\frac{\kappa}{a^2}\; .
\ee
From Eq.(5) it follows that if $\varrho +3p<0$ , $\ddot{a}$ will be positive, and the universe
will be accelerating.  For $p=\omega \varrho$ this means $\omega < -\frac{1}{3}$.
 We assume the relationship between pressure and energy density to be $p=-\varrho~(w=-1)$.

In this paper we consider the possibility that dark energy is the
Casimir energy due to noncommutative extra dimensions whose topology
is $S^2$.  A model in which dark energy arises due to the quantum
fluctuations of quantum fields in extra dimensions with simple
geometries has been carried out in \cite{mde}.  In that model scalar
fields contribute positive Casimir energy  only in the cases where
the number of extra dimensions is odd.  We show below that scalar
fields can contribute positive Casimir energy in even extra
dimensions and that the current experimental limits on the size of
the extra dimensions can be satisfied if the extra dimensions are
assumed to be noncommuting.

\section{The Non Commutative Sphere}
\par The noncommutative sphere is
defined in terms of three hermitian operators coordinates $x^{i}$ ($i=1,2,3$) satisfying the $SU(2)$ Lie algebra and an embedding constraint
 \be\label{comm}
\left[x^i,x^j\right]=-i\,\gamma\,{\epsilon^{ij}}_{k}x^k \; .
 \ee

 \be\label{embedding}
x^{i} x^{j} \delta_{ij}= a^2 \mathbf{1}
 \ee
where the noncommutative parameter $\gamma$, the radius  $a$ of $S^2$ and the dimension of the representation $M+1$ of $SU(2)$ are all related
by
 \be\label{costr}
 \label{uno}\frac{a}{\gamma} = \sqrt{\frac{M}{2}\Big(\frac{M}{2}+1}\Big).
 \ee
The last relation is obtained by computing the value of the
Casimir operator $x^i x^i= \gamma^2 \frac{M}{2} ( \frac{M}{2}+1)$
and setting this equal to $a^2$. For a fixed value of $a$, as the
value of $M+1$ increases (with the requirement that $M+1 \geq 2$),
$\gamma$ will decrease and according to (\ref{comm}) the sphere
becomes \textsl{less} noncommutative.
 $M+1=2$ represents the case of maximal noncommutativity, while for $ {M \rightarrow \infty}$ (or ${\gamma \rightarrow 0}$ ) we recover $S^{2}$.
The size of the representation $M+1$ assumes therefore a physical
meaning. The value of the Casimir energy to be interpreted as dark
energy, when NC is included, depends on both $a$ and $M+1$.

 To obtain the value of $a/\gamma$ given above we start from the $SU(2)$ Lie algebra
 \be
\left[L_i,L_j \right]= i {\epsilon_{ij}}^k L_k  \; .
 \ee
The total angular momentum squared commutes with everything and is
$L^2=L_iL_j \delta^{ij}$. The Hilbert space is spanned by
eigenvectors of the complete set of commuting observables $\{L^2,
L_3\}$:
 \be
L^2 |l,m>=l(l+1)|l,m> \quad L_3|l,m>=m|l,m> \; .
 \ee
The dimension of this Hilbert space is $2l+1$. We are calling this number $M+1$, therefore $l(l+1)$ becomes equal
to $ \frac{M}{2} (
\frac{M}{2}+1)$. Now we can make the change of notation $x^i/\gamma = L^i$ in order to put our algebra into the
standard $SU(2)$ form (i.e. eliminate the $\gamma$). Then by the above considerations of angular momentum we know that
$x^i x^i= \gamma^2 \frac{M}{2} ( \frac{M}{2}+1)\mathbf{1} $ which is equal to $a^2$ from the embedding constraint
(\ref{embedding}) and this proves (\ref{costr}).

\section{Computing the Casimir Energy for $\boldsymbol{M^4 \times S_{F}^2}$}
To compute the Casimir energy on  ${M^4 \times S_{F}^2}$ with $S_F^2$ being a noncommutative manifold, we modify the Green function method used
in \cite{milton1}.  Eq.(2.10) in \cite{milton1} is modified by truncating the  expression for the Green function which is expanded on
$S_{F}^{2}$ (see \cite{balac}) as
 \be \label{green} G_{M}(y,y')=\sum_{l=0}^{M}\sum_{m=-l}^{l} \frac{Y_{m}^{l}(y)\bar{Y}_{m}^{l}(y')}{l(l+1)+\mu^2}
\ee \label{green} where $y$ and $y'$ are the symbols of the operators defining $S^2_F$.  The effective mass shell condition is $\mu^2=
a^2(k^2-\omega^2)$.
\subsection{The Energy Density}
To obtain the energy density we first calculate the energy-momentum tensor of a massless scalar field defined
on $M^4 \times S^2$
 \be
  t_{AB}=\partial_{A}\varphi\partial_{B}\varphi-\frac{1}{2}g_{AB}\partial_{C}\varphi\partial^{C}\varphi\qquad
  (A,B=0\dots 5) \; .
   \ee
 The vacuum expectation value $<0|t^{00}|0>$ multiplied by only the volume of the two-sphere $V_{S^2} $ gives
 the Casimir energy on $S^2$ which is the energy density on our universe ($M^4$). Using the Green's function approach,
 the energy density for the fuzzy case is
\be
\label{u1} u(a)=-\frac{i}{(2\pi)^4}\int d^3 k \int_{c_{+}}dw\, w \sum_{l=0}^{M} \frac{2l+1}{\left(\frac{l(l+1)}
{a^2}+k^2-w^2\right)}
 \ee
where the only and key difference with Eq.(2.10) in~\cite{milton1} is that the summation goes to $M$ instead of infinity.
To derive this
expression we refer to the proprieties of the reduced Green's function in Eqs.(2.5) to (2.9) in \cite{milton1},
 and for the fuzzy sphere case we notice \cite{master} that
\be
 \sum_{m=-l}^{l} Y_{m}^{l}(y) \bar{Y}_{m}^{l}(y) = \frac{D_{l}}{V_{N}}=\frac{2l+1}{4\pi a^2}
 \ee
Eq. (\ref {u1}) can be written as
 \be
 \label{u2}
  u(a)=-\frac{ia^2}{(2\pi)^4}\int d^3k I(c_{+},\Sigma) \; ,
  \ee
 where
\be
\label{sum1}
 \Sigma(\omega)\equiv\sum_{m=1/2}^{M}\frac{m}{m^2-\beta^2}\\
 \ee
with
 \be
  m\equiv
l+\frac{1}{2}\qquad \beta^2=a^2\omega^2-a^2k^2+\frac{1}{4}\; .
\ee
With $m'=m+\frac{1}{2}$ the sum (\ref{sum1}) reads
 \be\label{exp}
 \Sigma(\omega)&=&\frac{1}{2}\sum_{m'=1}^{M}\Big[\frac{1}{m'+(\beta-\frac{1}{2})}-\frac{1}{m'}+\frac{1}
 {m'-(\beta+\frac{1}{2})}-\frac{1}{m'}+\frac{2}{m'}\Big]\nonumber\\
 &=&\psi \left( M+1/2+\beta \right) -\psi \left( M+1 \right) -\psi \left(1/2+\beta \right)-\gamma+\nonumber\\&&
 \psi \left( M+1/2-\beta \right)-\psi \left(M+1 \right) -\psi\left(1/2 -\beta \right) -\gamma+\nonumber\\&&
 2\,\psi \left( M+1 \right) +2\,\gamma\nonumber\\
 &=&\frac{1}{2}\big[-\psi(\frac{1}{2}+\beta)-\psi(\frac{1}{2}-\beta)+\psi(M+\frac{1}{2}+\beta)+\psi(M+\frac{1}{2}-\beta)\big].
 \ee
At this point our calculation starts to differ from Eq.(4.6) in \cite{milton1}. The last two terms are $M$-dependent
and are the result of the truncation of the sum in the expression for the Green's function.
 We consider any infinities contained in (\ref{u2}) as the contributions to the usual divergences of a scalar field
 on $M^4$, and we discard them. We now notice that the last two terms of Eq.(\ref{exp}) in the $\lim M \rightarrow \infty$
 give $-2\gamma+2\zeta(1)$. The integration of $ -\psi \left( 1/2+\beta \right) -\psi \left(
1/2-\beta \right) $ in (\ref{u2}) contributes to $\alpha_{KM}$, the energy density found in~\cite{milton1} for the
case $N=2$ only. We now calculate the contribution due to other two terms. We consider
 \be\label{psirep}
 \psi(\frac{1}{2}+M\pm\beta)=\ln(\frac{1}{2}+M\pm\beta)-\frac{1}{1+2M\pm 2\beta}-2\int_{0}^{\infty}\frac{t}{e^{2\pi
 t}-1}\Big[\frac{1}{t^2+[(1/2+M)\pm \beta]^2}\Big] \; .
 \ee
  In other words it is possible to follow \cite{milton1} replacing $\frac{(N-1)}{2}$ with $\frac{(2M+1)}{2}$,
  where $N$ is the dimension of $S^N$ and $M+1$ is the size of the representation of the fuzzy sphere. The $\Sigma(w)$ that
  we need to evaluate is now given by adding the six terms in (\ref{psirep}). The two logarithm terms can be written as
 \be\label{lnexp}
 -\ln\Big[\frac{2M+1}{2}\pm\beta\Big]=-\frac{1}{2}\ln\Big[\Big(\frac{2M+1}{2}\Big)^2-\beta^2\Big]+\frac{1}{2}\ln\Big[\frac{(2M+1)/2\mp\beta}{(2M+1)/2 \pm\beta}\Big]
 \ee
 The second two terms of (\ref{lnexp}) must be added (see Eqs.(16) and (17)) and give zero. We write the first term as
   \be\label{ulog}
  -\frac{1}{2}\ln\Big[\Big(\frac{2M+1}{2}\Big)^2-\frac{1}{4}+a^2 k^2-a^2 \omega^2\Big]=
  -\frac{1}{2}\ln\Big[\Big(C+a^2 k^2-a^2 \omega^2\Big] \; .
  \ee
  with
  \be\label{Cdef}
  C=\Big(\frac{2M+1}{2}\Big)^2-\frac{1}{4}
  \ee
The presence of the constant C gives the contribution to the
energy density $\alpha_{C}$. Utilizing equations (2.14) to (2.16),
in \cite{milton2}, we obtain
 \be\label{alphaC}
 \alpha_{C}=-\frac{1}{2^5\pi ^2}\int_{0}^{-C} dz'' \, \int_{0}^{z''} dz' \, \int_{0}^{z'} dz
 \frac{D'_{\beta}}{\beta}|_{\beta=z+1/4}=\frac{C^3}{96\pi ^2}
 \ee
 For $M=1$, which means size of the representation is $2$, the numerical value of (\ref{alphaC}) is
 \be
\alpha_{C}=\frac{1}{12\pi ^2}
  \ee
 Although the we have followed the procedure of \cite{milton2} the physical meaning of $C$ in our case is different from
 that of \cite{milton2}.
 In \cite{milton2} $C$ depends on both $N$
 and the mass of the fields, while in our case it depends only on $M$. We consider a scalar massless field on $S^2$ only,
 therefore the linear and quadratic terms in $c$ of Eq.(2.25) in \cite{milton2} are null.
 Our definition of $\beta$ is different from \cite{milton2} since it does
 not include the constant $c$.
 The evaluation of the remaining four terms of (\ref{psirep}) can be done based on the algebraic replacement of
 $\frac{(N-1)}{2}$ by $\frac{(2M+1)}{2}$. Keeping the same argument about the contours of integrations,
 the position of the poles switch to
\be\label{pole}
 \beta=\pm (1+2M)/2\qquad \rm{and} \qquad \beta=\pm(1+2M)/2\pm it
 \; ,
\ee
 with the result
 \be\label{alphaNC}
 \alpha_{NC}=\frac{1}{16\pi ^2}\int_{0}^{\infty}\frac{dt}{e^{2\pi t}-1}Im\{D_{it}[i(2M+1)t-t^2]^2\}
 \ee
For $M=1$ the numerical value of (\ref{alphaNC}) is
 \be
\alpha_{NC}=-\frac{0.0060}{\pi ^2}  \; .
  \ee
Using $\alpha_{KM}=-\frac{1}{1260\pi}$ we can evaluate the energy
density (\ref{uc1})
 \be
 \label{alphaTOT}
\alpha_{TOT}=\alpha_{KM}+\alpha_{NC}+\alpha_{C}=0.0077 \; .
 \ee
 We notice that the first two terms are negative, while the third one is positive, and its contribution makes the
 overall sign of the energy density positive.
\subsection{The Commutative Limit}
The commutative limit is given by $\lim {M \rightarrow \infty}$ of (\ref{alphaTOT}). We take this limit in Eq.(\ref{exp}),
and we obtain
 \be
\lim_{M\rightarrow \infty} \psi(M+\frac{1}{2}+\beta)+\psi(M+\frac{1}{2}-\beta)=-2\gamma+2\zeta(1) \; ,
 \ee
 which is infinite. We can consider it to be the contribution to the usual vacuum energy density present on $M^4$
 and discard it. For $M\neq
 \infty$ a finite contribution to the energy density is given by $\alpha_{NC}+\alpha_{C}$. Choosing a manifold such as
  $M^4 \times S^2$ allows
 us to regularize the infinities present on $S^2$. A similar problem was encountered in \cite{demetrian} where the
 Casimir energy for only $S_{F}^2$ has been computed. However in \cite{demetrian} a finite value was not recovered
 in the commutative case.
\section{Cosmological Interpretation Of The Result}
In order for our model to be a viable description of dark energy, the required size of the extra dimensions must be consistent with current
experimental limits on the size of extra dimensions. To estimate the size of the extra dimensions obtained from our model, we can set the energy
density $u_{C}$ equal to the observed value and estimate the value of dark energy density present in our universe, $u_{C}=\frac{3H_{0}^2}{8\pi
G}=1.05\times 10^{-5} h_{0}^{2} \rm{GeV cm^{-3}}$ (with $c=\hbar=1$ and conversion factor $\hbar c=2\times 10^{-14}$ GeV cm). We can then
numerically solve for $a$ in Eq.(\ref{U}), and we obtain
 \be\label{avalue}
 a=\alpha_{TOT}\Big[\ln(\frac{a}{L_{Pl}})\Big]^\frac{1}{4} 80\mu m=71.92\mu m
 \; .
 \ee
 The value of $a=71.92\mu m$ is not ruled out by tests of classical gravity at small distances, which indicate no deviation down to $200\mu
 m$. Our model is only a toy model since the number of extra dimensions is unknown. String theory for example requires 6
 extra dimensions while
$M$ theory requires 6+1. Hence it becomes obvious to look for the values of $\alpha_{TOT}$ for other values of $N$ and other sizes of the
representation $M+1$ as well. We notice from Eqs.(33) in \cite{milton1}, (\ref{alphaC}) and (\ref{alphaNC}) that a larger $N$ contributes to a
more negative value of $\alpha_{KM}$. A larger $M$ will provide a more negative contribution of $\alpha_{NC}$ and a more positive (and dominant)
contribution for $\alpha_{C}$. Therefore there are additional solutions which allow a positive overall sign for $\alpha_{TOT}$. In order to
provide the right value of the dark energy density (\ref{U}), as $\alpha_{TOT}$ increases $a$ must increase, but $a$ must stay below the
experimental limit of $200\mu m$. So it seems possible to obtain a suitable value of $\alpha_{TOT}$ for larger $N$ and $M$ combined. The latter
analysis will be carried out in a subsequent paper. This model  also allows us to impose numerical limits on intrinsic NC features, such as
$M+1$, on a cosmological level.
\
\end{document}